\documentclass{emulateapj}
\shorttitle{XMM-Newton observations of Mrk 110}
\shortauthors{Dasgupta \& Rao}
\usepackage{graphicx}
\journalinfo{Astrophysical Journal Letters}
\submitted{}
\begin{document}
\title{Discovery of hard X-ray delays in the X-ray emission of the Seyfert 1 galaxy 
Mrk~110: possible evidence for Comptonization }
\author{Surajit Dasgupta \altaffilmark{1}}
\author{A. R. Rao \altaffilmark{1}}
\altaffiltext{1}{Tata Institute of Fundamental Research, Mumbai-400005, India,
email: surajit@tifr.res.in, arrao@tifr.res.in, url: http://www.tifr.res.in/$\sim$surajit}



\begin{abstract}
We report the discovery of  hard X-ray delays in the  X-ray emission of the 
Seyfert 1 galaxy Mrk~110, based on a long XMM-Newton observation. Cross 
correlation between the X-ray light curves of different energy bands reveals 
an energy dependent delay ranging from a few minutes to an hour. We find that 
the energy spectrum can be modeled by Comptonization of disk blackbody photons. 
The energy dependent delay can be modeled as due to the effect of Comptonization 
in a hot plasma confined within 10 Schwarzschild radius of the black hole. We 
discuss our results in the context of inverse Comptonization of the soft photons 
by highly energetic plasma.
\end{abstract}

\keywords{galaxies: active --- galaxies: individual (Mrk 110) --- radiation 
mechanisms: general --- accretion, accretion discs --- X-rays: galaxies}

\section{Introduction}
The dominant X-ray radiation mechanism in accreting black holes is 
commonly thought to be inverse Compton scattering of low energy 
photons by a cloud of hot ($T\sim 10^{8} - 10^{9}~K$) electron plasma 
near the black hole \citep{st80}. The physical geometry 
of the coronal plasma responsible for scattering the photons is not well 
constrained by the current observations. The ubiquity of a 
disk blackbody component accompanied by a power-law tail in the overall 
spectra of galactic black hole candidates (GBHCs) motivated several 
workers to develop the so called two phase accretion disk-corona models 
\citep[see eg.,][]{hm91,hm93, pkr97,sz94,ste95,bel98}. In these models, 
the blackbody radiation from the cold disk enter the hot corona and is
Comptonized into X-rays. A part of the hard X-rays from the corona, being 
reprocessed in the disk, produce the reflection hump. The geometry 
of the corona controls this feedback mechanism which in turn determines 
the spectral slope of the escaping radiation. The Kompaneets y-parameter 
(and hence the temperature) is determined by the energy balance between 
heating and cooling mechanism inside the plasma. One important question 
in all such models is the method by which the gravitational energy is 
converted to the energy of electrons. The ideas explored in the literature 
include magnetic flares \citep{hmg94,ste95,pou96,bel99a,bel99b}, advection 
dominated disk very close to the black hole \citep{ny94}, Bondi type 
free fall beyond a shock region \citep{ct95}, but, as yet, there is no 
consensus on the exact mechanism. A general prediction of the above 
models is that the hard X-ray variations should lag behind those in 
softer bands, as hard photons undergo more number of scatterings in the 
plasma before escape. Measurement of such time-lags has the potential of 
constraining the size of the region which in turn will help us to understand 
the mechanism by which the electron cloud is energized.

Frequency dependent time lags have been observed in Galactic black hole 
candidates \citep{miy88,mkk91} as well as in a few active galactic nuclei 
\citep[AGNs;][]{pnk01,zh02}. Interpreting these lags as due to Comptonization 
requires the size of the emitting region to be very large, typically several 
thousand Schwarzschild radii (compatible to the lowest frequency at which the 
lag is observed), leading to the problem of heating the electron cloud at large
distances from the black hole. Hence, sometimes these lags are interpreted as
due to the energy dependent asymmetries in random shots \citep{mk89}; that 
is interpreting the lags as due to the production of the variability itself.

One of the problems of detecting lags at higher frequencies and relating 
them to the Comptonization at the innermost regions of the accretion disk 
around black holes could be observational. If the Comptonization process
occurs at 10 -- 20 Schwarzschild radii and if the Comptonization process
gives a lag which is a factor of few larger than the light travel time in 
these regions, one expects a lag of a few tens of milliseconds in a black 
hole of mass 10 M$_\odot$ and  about a day in a AGN of mass 10$^8$ M$_\odot$. 
Detection of such delays is observationally a difficult task. Bright nearby 
AGNs with a black hole mass of 10$^7$ M$_\odot$, where one expects a delay 
of about an hour, are the ideal sources to look for delays due to the process 
of Comptonization. With this motivation, we have searched for delays in 
one of the bright low mass AGN Mrk~110, based on a long observation using 
the XMM-Newton observatory. In this {\it Letter} we present the cross-correlation 
analysis in different energy bands of this source and show that hard X-rays 
are delayed by a significant amount of time (from few minutes to an hour). 
The 0.3 - 12 keV X-ray spectrum can be represented by a  Comptonization 
model which can explain the measured hard X-ray delay.

\begin{figure}
\centering
\includegraphics[scale=0.35,angle=-90]{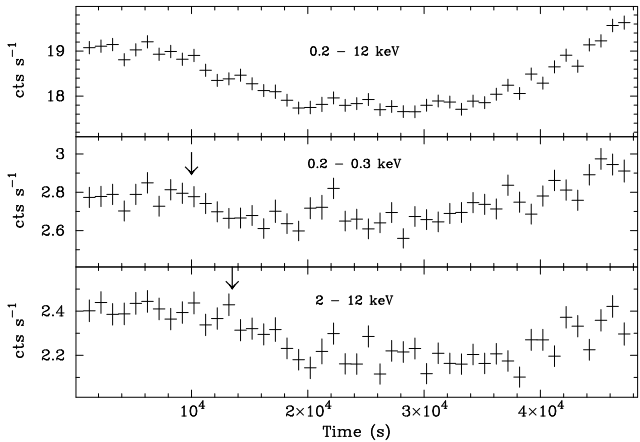}
\caption{EPIC-PN light curves (1000 s bin) of Mrk~110 in different energy
range. Start of variability is indicated by downward arrows (see text)}.
\label{lc}
\end{figure}

\section{Observation}
Mrk~110 is a nearby optically bright, radio intermediate ($R\sim1.6$) 
narrow-line Seyfert 1 galaxy (NLS1s) at a redshift z=0.036. The optical 
continuum and the broad emission lines of Mrk 110 are highly variable 
\citep[by a factor of 2 to 8 within a timescale of 10 years - ][]{kw01,kol03}. 
Mrk~110 was observed on 2004 November 15 by XMM-Newton for 47.4 ks. The 
EPIC-PN cameras were operated in Prime-Small-Window observing mode using 
the Thin1 filters. Here we use data only from EPIC-PN cameras due to the 
better efficiency, better calibration at lower energies and absence of 
pile-up. Source spectra and light curves were extracted from the EPIC 
images using a circular source region centered on the observed source 
position. Background spectra and light curves were derived from adjacent 
`blank sky' regions. The EPIC spectra were binned to give a minimum of 
50 counts per bin. The {\sc xspec v11.0} and {\sc xronos} packages 
were used for spectral and timing analysis respectively. Errors on fitted 
parameters are quoted at the nominal 90\% confidence level ($\Delta \chi^2$ 
= 2.7) unless otherwise stated.

\section{Timing Analysis \label{timing}}
In Figure~\ref{lc} we plot the binned light curves of Mrk~110 in different 
energy ranges (not corrected for 71\% duty cycle of PN SW mode). The $0.2 -12$ 
keV light curve shows peak to peak variation of approximately 10\% within 3 
hours. To quantify the source 
variability in different energy bands, we calculated the fractional variability 
in seven energy bands: E$_1$ (0.2 -- 0.3 keV), E$_2$ (0.3 -- 0.42 keV), E$_3$ 
(0.42 -- 0.58 keV), E$_4$ (0.58 -- 0.8 keV), E$_5$ (0.8 -- 1.2 keV), E$_6$ 
(1.2 -- 2 keV), and E$_7$ (2 -- 12 keV), respectively. The energy bands are 
chosen such that the mean count rate in those bands are approximately same 
(within 10\% of average). We find significant variability (rms value  2.5-4.5\%) 
in all the energy bands for 500 s binning (with a typical error of 0.4\%). 
There is a marginal evidence for increasing variability with increasing energy: 
the variability is (2.7 $\pm$ 0.4)\% for energy $<$0.6 keV and it is (3.6 $\pm$ 
0.4)\% for energy $>$0.6 keV. A Structure Function analysis \citep{rut78} 
of the light curves shows that the shortest correlation timescale is more than 
a few thousand seconds. 


\begin{figure}
\centering
\includegraphics[scale=0.5,angle=0]{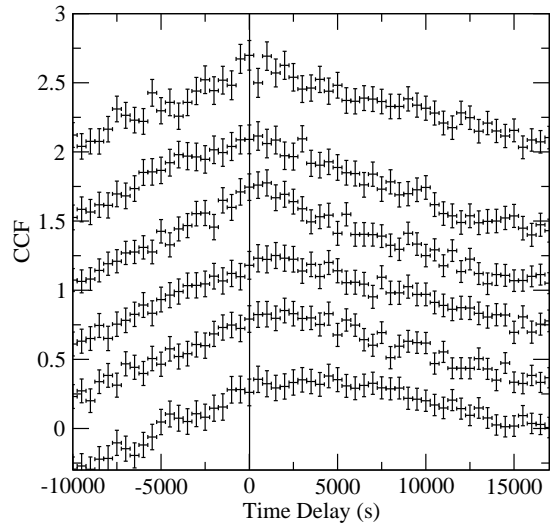}
\caption{The CCFs between light curves of 0.2 -- 0.3 keV and higher energy bands 
(increasing from top to bottom). The CCF values are vertically shifted for clarity. 
(see text)}
\label{ccf}
\end{figure}

\begin{figure}
\centering
\includegraphics[height=8.8cm,angle=-90]{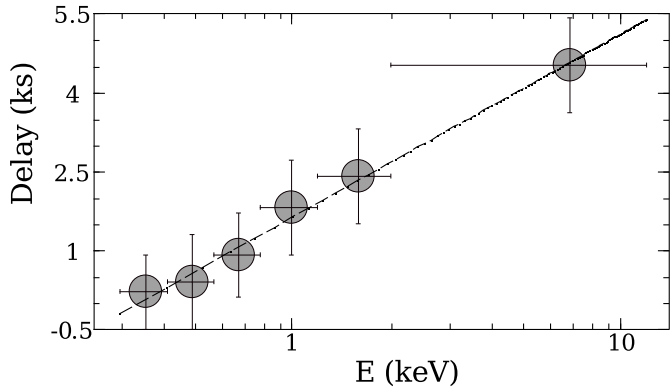}
\caption{Energy dependence of lag between the light curves of energy 0.2 -- 0.3 keV 
and higher energies.}
\label{delay}
\end{figure}

A visual examination of the light curves reveals a gradual decrease in count 
rate ($\sim 0.2$ cts in 2 hrs) in E$_1$ at time 10 ks from the starting time. 
Similar kind of variation is seen after more than an hour in E$_7$. (marked 
by arrows in figure~\ref{lc}). To search for time lags, we calculated the 
cross correlation function (CCF) between E$_1$ and other energy bands using 
the {\it crosscor} package in {\sc ftools}. The CCFs between E$_1$ and E$_i$
(where i=2,3,..,7) are plotted as a function of delay, the successive plots 
are vertically shifted by 1.6, 1.2, 0.9, 0.6, 0.3, 0 respectively for clarity.
The errors in the CCF values are the standard one sigma values which includes 
only the counting statistics errors. The harder light curves systematically 
lag the softer band. To estimate amount of the delay, we fitted the central 
part of the CCF distribution with a Gaussian function and derive values of 
delays for the different energy bands of 200$\pm$700 s, 400$\pm$900 s, 
900$\pm$800 s, 1800$\pm$900 s, 2400$\pm$900 s, 4500$\pm$900 s, respectively. 
The errors in the delays are estimated by the $\chi^2$ fitting method with the 
prescription $\Delta \chi^2$ = 4 (for three parameters). In Fig~\ref{delay} 
we plot the derived delay as a function of the energy of the hard band, along with a 
Comptonization model (see section~5) fitted to the data points ($\chi^2$/dof 
$\sim$ 1/4). For the hypothesis that there is no delay we get a value for $\chi^2$/dof 
of 38/6 and for the hypothesis of energy independent delay gives a value for 
$\chi^2$/dof of 18/5. Hence we can conclude that a delay is detected and it is 
energy dependent at more than 99\% confidence level. We must, however, 
caution  that the above results are based on statistical considerations only and 
do not include systematics like the shape of the variation of CCF with delay. We 
have also derived cross correlation using other combinations (CCFs of $E_i$, w.r.t. 
$E_4$, where i=1,2,3,5,6,7) to check whether the results are artifact of uncertainty of instrumental calibration of energy below 0.3 keV and we find similar results.

\section{Spectral Analysis \label{spec}}
We first modelled the X-ray spectrum of Mrk~110 with a absorbed power-law 
with Galactic column density $N_H=1.6 \times 10^{20}$. This provides a fairly 
good fit to the PN data in the 2--12 keV range ($\chi^2/\nu = 1037/1022$) 
giving a photon index $\Gamma = 1.75\pm0.01$.Inclusion of a narrow Gaussian 
emission line improves the fit significantly 
($\chi^2 / \nu = 997/1020$), giving the energy of the line  $6.42\pm0.02$ keV.
Extrapolating the best fit model to 0.3 keV shows a huge soft excess in the 
spectrum ($\chi^2/\nu = 94471/1363$). Refitting still provides a rather poor 
fit ($\chi^2/\nu = 7350/1363$). A power-law plus a blackbody (generally the soft 
excess component in AGN is modeled by thermal black-body but the origin is still 
unknown) to model the soft excess improves the quality of fit but it is still 
not acceptable ($\chi^2/\nu = 2109/1361$). 
The temperature of the blackbody becomes $kT=100\pm2$ eV. Fitting the data 
with more realistic thermal accretion disc spectrum like {\it diskbb} 
\citep{mit84} or {\it diskpn} \citep{gier99} provides a poor fit to
data ($\chi^2/\nu = 1780/1361$ and $\chi^2/\nu = 1782/1360$ respectively)
and the temperature becomes very high ($kT = 155\pm2$ eV and  $159\pm3$ eV 
respectively). These results would seem to reject an origin for the 
soft excess in terms of unmodified thermal blackbody emission. Addition of 
a power-law instead of blackbody improves the fit ($\chi^2/\nu = 1680/1361$). 
The values of the power-law indices are $2.47\pm0.02$ and $1.21\pm0.04$. A 
model in 0.3 -- 12 keV range which consists of a broken power-law and 6.4 keV 
Gaussian emission line improves the fit slightly ($\chi^2/\nu = 1597/1361$). 
The power-law indices are $2.29\pm0.01$ and $1.78\pm0.01$, and the break energy 
is $1.66\pm0.04$ keV. 

Reflection off the surface could also produce strong soft excess at low 
energies. To test this the ionized reflection model {\it pexriv} \citep{mz95} 
was fitted to the 0.3 - 12 keV spectrum resulting in a poor fit to the data 
($\chi^2/dof \sim 2185/1361$). The best fitting parameters were 
$\Gamma = 2.26\pm0.01$, $R=8.14\pm0.12$ and $\xi < 10^{-3}$.

The energy dependent time lags are strongly suggestive of Comptonization
of low energy seed photons by a population of high temperature electrons 
(section~1). The {\sc comptt} code \citep{tit94} was used to 
model Comptonization of soft photons in a thermal plasma. A power-law plus 
a Comptonization component give a good fit to the data ($\chi^2/\nu = 1477/1359$) 
with $\Gamma = 1.51\pm0.05$ and the seed photon temperature $kT_{bb} = 69\pm2$ 
eV. The temperature and optical depth of the Comptonizing plasma are strongly 
covariant parameters, and thus cannot be constrained simultaneously. Fitting
the data with the {\it compps} model \citep{pou96} along with the power-law 
gives a reasonable fit ($\chi^2/dof = 1483/1359$). In this model the soft 
flux is equal to the sum of the absorbed incident flux from the corona and the 
flux due to local energy dissipation in the cold disk. The spectral shape of 
the  soft components are assumed to be Planckian, with temperature $T_{bb}$ 
and $T_{disk}$, respectively ($T_{bb}>T_{disk}$). The inner disc temperature 
$kT_{disk}$ is fixed at $40$ eV (calculated for black hole of mass $10^7~M_{\odot}$ 
assuming standard accretion disk). The best fit values of the parameters are 
$\tau = 3.2^{+1.6}_{-0.7}$, $kT_e = 14.0^{+3.1}_{-3.2}$ keV, 
$kT_{bb} = 99^{+5}_{-5}$ eV, $\Gamma = 1.57^{+0.04}_{-0.05}$.

\begin{figure}
\centering
\includegraphics[height=8.8cm, angle=-90]{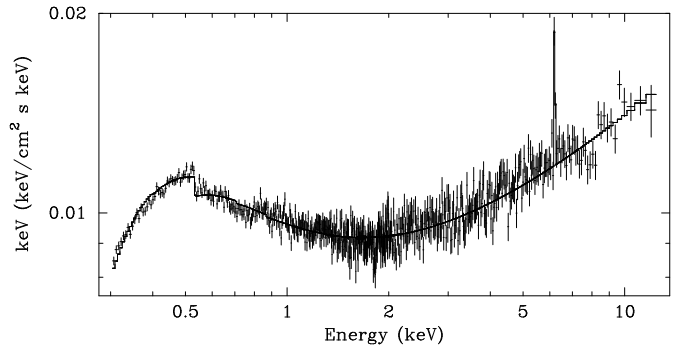}
\caption{XMM-Newton (EPIC-PN) spectrum in the energy range 0.3 -- 12 keV and the 
best fit hybrid thermal/non-thermal Comptonization model.}
\label{spectrum}
\end{figure}

The Comptonization model along with a power-law described above has the two
spectrally identified continua originating in two distinct thermal Comptonizing
plasmas. An alternative is that the whole spectrum is produced by a single
plasma with a hybrid thermal/non-thermal electron distribution \citep{cop99}.
The definite detection of a non-thermal Comptonization component requires high 
energy and high quality data. As noted by \citet{pdo95}, the X-ray spectra of 
ultrasoft Seyfert galaxies do resemble that of Cyg X-1 in its high/soft state. 
Thus it seems reasonable to test whether a hybrid thermal/non-thermal plasma 
is a viable model for the X-ray continuum for Mrk~110. To test this  idea we 
tried the hybrid Comptonization model {\it compps} \citep[for a detailed 
description of parameters see][]{zd05}. We assume a spherical geometry of hot 
plasma. We fitted the model ($\chi^2/dof = 1455/1359$) and the best fit values 
of the parameters are $\tau = 4.8^{+1.6}_{-2.1}$, $kT_e = 11.1^{+9.4}_{-1.8}$ keV, 
$kT_{bb} = 105^{+4}_{-3}$ eV, $\Gamma_{inj} = 2.43^{+0.09}_{-0.12}$, 
$\gamma_{min} = 1.19^{+0.13}_{-0.05}$. The unfolded spectrum is shown in 
Figure~\ref{spectrum}. Similar attempt has been done to model the X-ray spectrum 
of NLS1 galaxy Ton S180 \citep{vau02}. 

\section{Discussion and Conclusion}
We find significant energy dependent delay between the hard and soft X-ray
emission in the Seyfert 1 galaxy Mrk 110. Similar delays were also found in the
bright Seyfert 1 galaxy MCG-6-30-15 \citep{pon04}. \citet{gal04} found alternating 
leads and lags in the NLS1 IRAS 13224-3809. Though other interpretations like 
geometric effects and energy dependent shape of shots can also explain the 
observed energy dependent delays in Mrk 110, we try to interpret the results 
as due to Comptonization, particularly because of the fact that the spectral 
analysis favours a two component Comptonization model. We consider a 
static Compton cloud with optical depth $\tau_T$, and electron temperature 
$\Theta=kT_e/m_e c^2$. A soft seed photon of energy $E_0$ injected into the 
cloud increases its energy by a factor of $A=1+4\Theta+16\Theta^2$, on an average
after each scattering, so that after n scattering its energy is $E_n=A^n E_0$.
The photon mean free path is $\lambda \approx R/max(1,\tau_T)$ (where R is the 
size of the X-ray emitting region). The time difference between successive 
scatterings is then $t_c = (R/c)/max(1,\tau_T)$, where $\tau_T= N \sigma_T R$ is 
the Thompson optical depth, and $\sigma_T$ is the cross section of Thompson 
scattering, N is the electron number density of the scattering medium. The time 
needed to reach the energy $E_n$ is $t_n=nt_c$ \citep{st80,pay80}. We have 
calculated time lags of different high energy bands with respect to 0.2 -- 0.3 
keV band. Hence in our case $E_0=0.25$ keV.
We fit the above equation ($t_n=nt_c$) to the result (hard lag as a function 
of energy) of our analysis and find that the data is in good agreement with the 
equation (Fig.~\ref{delay}). Using the values of the best fit parameters of the 
above result and the parameters of the Comptonization model fitted to spectrum 
we get $R\sim2\times10^{11}~m~\sim 10~R_S$. This value of R is physically realistic 
as most of the energy is dissipated within 10 $R_S$. \citet{bl98} pointed out 
that any scenario in which the observed hard time lags are purely due to static 
Comptonization requires that the radial extent of the hot corona exceeds 
$\sim 10^{4}~R_S$ of a solar mass black hole ($R>10^{8}~m$). This is incompatible 
with current models of accretion flows onto galactic black holes \citep[see 
the review by][]{lia98} even from simple energy arguments (see section~1). 
But the results reported here pertains to lags at very short times scales and 
hence is consistent with the static Comptonization model. The harder power-law 
emission extending to 12 keV (and presumably beyond) can also be produced by 
Comptonization, either in another purely thermal plasma or non-thermal electrons 
in a plasma with a hybrid thermal/non-thermal distribution. 

The origin of the hot plasma above the accretion disk is quite debatable. If the 
radiation pressure inside the accretion disk very close to the black hole is very 
high then the accretion disk can be puffed up and the hot thermal plasma can be
formed just above the accretion disk. Hybrid thermal/non-thermal plasmas 
have often been successfully used to model the observed data \citep{gier99,pc98}.
A possible origin of the non-thermal component is in process of magnetohydrodynamic 
turbulence occurring in the corona \citep{lm97}. Stochastic acceleration of 
particles by stochastic gyroresonent acceleration in accreting plasma can also 
accelerate particles to higher energies \citep{dml96,lkl96}. 

We conclude that inverse Compton scattering of soft photons by highly energetic 
electron distribution provides a satisfactory explanation of the hard X-ray time 
lag observed in the XMM-Newton observation. The energy spectrum in the 0.3 -- 12 
keV band can be modelled either by two component Comptonization or by a hybrid 
thermal/non-thermal Comptonization. It is not possible to distinguish between 
them without good quality high energy data.

This work is based on observations obtained with the XMM-Newton, an ESA science 
mission with instruments and contributions directly funded by ESA member states 
and the USA (NASA). Authors are grateful to H. Netzer, G. Dewangan and S. Mandal 
for useful suggestions. We thank the referee for useful comments.

\end{document}